\documentclass[epj-spec]{svjour}
\usepackage{graphics,color}
\begin{document}
\title{Optical Scar in a chaotic fiber}
\author{Val\'erie Doya\thanks{\email{valerie.doya@unice.fr}} \and Olivier Legrand \and Claire Michel \and Fabrice Mortessagne }
\institute{Laboratoire de Physique de la Mati\`ere Condens\'ee, Universit\'e de Nice, France}
\abstract{We propose to use a multimode optical fiber with a D-shaped cross section as a privileged system to image wavefunctions of a chaotic system. Scar modes are in particular the subject of our investigations. We study their imprints on the statistics of intensity and we show how the introduction of a localized gain region in the fiber is used to perform a selective excitation of scar modes. 
} 
\maketitle
\section{Introduction}
\label{intro}
Experimental wave chaos has naturally emerged from the demand to achieve simple experiments to study quantum chaos. Since 1980, a lot of experiments have been performed in different systems and with different natures of waves. Some years ago, as we were interested in the study of the spatial properties of waves that propagate in a chaotic medium, we asked the following question: how to image  wave chaos experimentally? In the litterature, some nice ways to vizualize the wavefunctions of chaotic billiards have already been proposed. An exhaustive presentation of wave chaos experiments is reported in the St\"ockmann's book\cite{Stockmannbook}, we briefly cite the main experiments related to wave chaos imaging. Spatial distributions of intensity have first been observed in microwave cavities by Sridhar \textit{et al}\cite{Sridhar} showing a selection of wavefunctions associated with some particular periodic orbits. By measuring the field displacement of vibrating plates, Schaadt \textit{et al.}\cite{Schaadt} have investigated the statistical properties of the wavefunctions of a Sina\"i stadium plate. In the afore mentioned studies, the measurements were nondirect: the mapping of the wavefunctions was obtained by performing a point-to-point scan of the surface (of the microwave cavity, or of the aluminium plates).\\
A direct visualisation of the wave patterns of a stadium has been achieved on the surface of water enclosed in a stadium-shaped container by Bl\"umel \textit{et al.}\cite{Blumel}. This type of experiment prevents any quantitative analysis of the statistical properties of the field, but has produced interesting results on the direct observation of scarred patterns\cite{Kudrolli}.\\ Optical waves also constitute an evident candidate to image wave chaos. One of the first experiment with optical waves invoking the properties of classical chaos to explain the behavior of waves has been proposed by the group of A. D. Stone\cite{Stone1}. It has been shown that the lasing properties of deformed droplets can be understood \textit{via} a transition to chaotic ray dynamics. Until 2002, no direct observation of the wavefunctions of a chaotic optical billiard had been realized. With the aim of studying the spatial properties of waves in a chaotic system, we have then proposed an experiment of wave chaos in an optical fiber. The experiment, simple in its concept, offers a direct access to the spatial distribution of intensity by imaging the near-field intensity at the fiber output. Unlike previous experiments, this experiment also gives the opportunity to investigate the properties of a field being the result of the propagation of a superposition of several modes. 
In this paper, we present an overview of our recent results on the spatial properties of the wavefunctions of a chaotic multimode optical fiber. From the experimental validation of Berry's conjecture on the generic ergodic behavior of wavefunctions of a chaotic billiard\cite{Berry1} to the direct observation of a scar mode, optical fibers provide a good experimental system to image wave chaos.\\ 
This paper is organized as follows. In section \ref{system}, the experimental setup is described and its relevance in the field of wave chaos is presented. Our main results about the experimental observation of the typical behavior of wavefunctions in a chaotic billiard are reported in section \ref{TypResults}. In order to investigate the \textit{scar} wavefunctions, we have developped two approaches: the first one is dedicated to the signature of a scar wavefunction on the statistics of the field and intensity, the other one to a selective excitation of scar by the way of optical amplification. Theses recent studies are described in section  \ref{Scarsection}. 

\section{Presentation of the experimental system}
\label{system}
\subsection{Description of the system}
\label{fiber}
\begin{figure}[ht]
\input{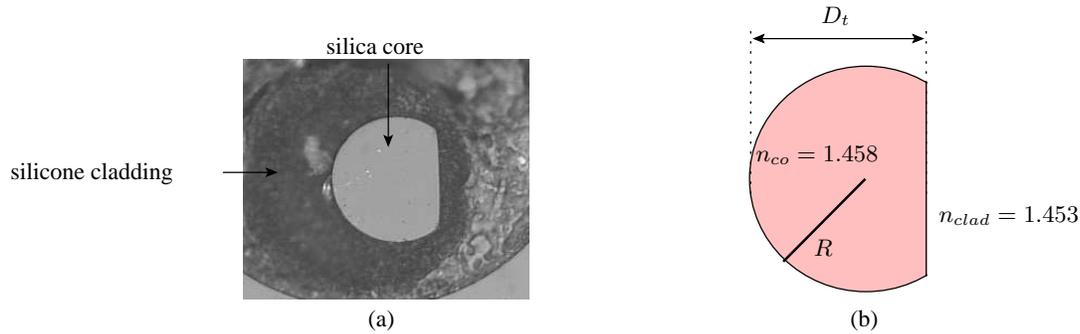}
\caption{Microscope photography of the D-shaped fiber(a) and characteristical parameters of the core(b).}
\label{figfiber}       
\end{figure}

\begin{figure}[h]
\centering
\input{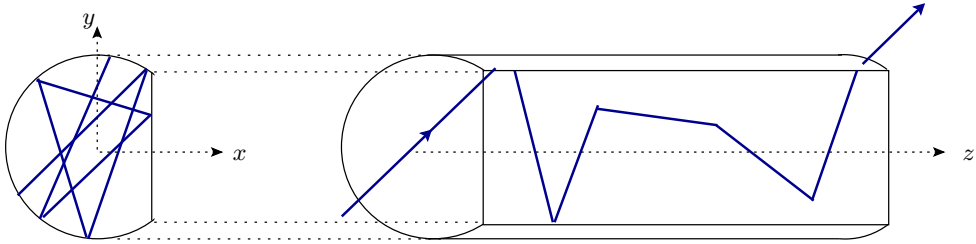}
\caption{Light propagation along the fiber in the geometrical limit of rays.}
\label{figbilliard}       
\end{figure}

The multimode optical fiber designed and fabricated in our lab for our studies is briefly described in this section. The transverse cross section of the fiber is a truncated disk fabricated from a 1-cm-diameter preform of silica that has been cut at half its radius and pulled at a temperature low enough to maintain the D-shaped cross section. The final dimensions of the fiber after pulling are 120 $\mu$m for the diameter and 90 $\mu$m for the truncated diameter (see Fig. \ref{figfiber}). These dimensions differ from the typical 10-$\mu$m-diameter of a single mode optical fiber used in the field of optical telecommunication. The index of refraction of the silica core $n_{co}$ at our working wavelength is around 1.458. It is surrounded by a black absorbing silicon cladding with an index of refraction equal to $n_{clad}$=1.453. The cladding is tinted in order to prevent light from being guided during the propagation.\\
\\
As the characteristic dimensions of the fiber are large compared to the optical wavelength, a collimated light beam  that propagates along the fiber can be approximated, in the geometrical limit, to a ray that is reflected at the core/cladding interface along the propagation in the fiber. A formal derivation of the eikonal equation from the 3D Helmholtz equation is given in \cite{Doya2} and justifies the analogy of our system with a chaotic billiard. Indeed, the evolution of a ray along the fiber (in the direction of $z$) can be associated to a time evolution on the transverse cross section of the fiber (see Fig. \ref{figbilliard}).
\\
The so-called D-shaped billiard has been proved to be a fully chaotic billiard by Bunimovich\cite{Bunimovich} and Ree\cite{Ree}, as soon as the truncated diameter $D_t$ is larger than half the radius. If the optical fiber can be seen as a wave billiard, it also constitutes an interesting analogous quantum system. Indeed, the propagation of the field along the fiber is described by the Helmholtz equation. In the weak guidance regime, i.e. for a small difference of indices between the core and the cladding $[(n_{co}-n_{clad})/n_{clad}\ll 1]$, a given polarization is nearly perfectly preserved along the propagation thus enabling to use the scalar form of the Helmholtz equation:
\begin{equation}
(\Delta+\partial_{zz})\psi(\textbf{r},z)+n(\textbf{r})^2k_0^2\psi(\textbf{r},z)=0
\end{equation}
with $k_0$ the vacuum wavenumber and $n(\textbf{r})$ the index of refraction in the medium. In the paraxial approximation, this equation reduces to the following form:
\begin{equation}
i\beta_{co}\partial_z\phi(\textbf{r},z)=[-\frac{1}{2}\Delta+V(\textbf{r})]\phi(\textbf{r},z)\label{Helmholtz2}
\end{equation}
where $\beta_{co}=n_{co}k_0$ and $\phi(\textbf{r},z)$ is obtained from $\psi(\textbf{r},z)=\phi(\textbf{r},z)e^{i\beta_{co}z}$. 
The paraxial approximation is associated to the propagation of light close to the optical axis and is fulfilled as soon as the angle of the guided ray is much smaller than the cutoff angle $\theta_{max}$ which is given by $\sin\theta_{max}=\sqrt{1-(n_{clad}/n_{co})^2}$. \\
Equation \ref{Helmholtz2} is then formally equivalent to a pseudo-time independent Schr\"odinger equation. As a consequence, the D-shaped fiber appears to be an ideal wave chaos paradigm of a quantum chaos experiment. 

\subsection{Typical results}
\label{TypResults}
\begin{figure}[h]
\centering
\input{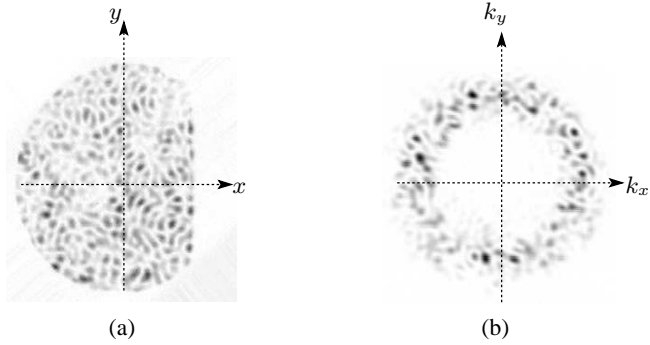} 
\caption{Near-field intensity (a) and far-field intensity (b) for $\kappa_t=36/R$.}
\label{Speckle}       
\end{figure}

\begin{figure}[h]
\centering
\input{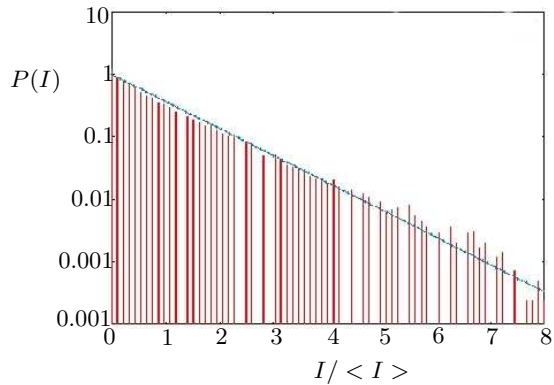} 
\caption{Probabiliy distribution of the intensity associated to the figure \ref{Speckle}(a). The Poisson's distribution prediction is plotted in dotted line.}
\label{PT}     
\end{figure}

One advantage of the experiment with the optical fiber is that the experimental setup is conceptually simple. A beam from a Helium-Neon laser source at $\lambda_{He/Ne}=628$ nm is collimated and enlarged so that its size becomes large enough compared to the transverse size of the fiber to be in a quasi-plane wave illumination. The main parameter of the experiment is then the angle $\theta$ between the beam with the optical axis at the fiber entrance. Indeed, this angle governs the value of the transverse wave number in the fiber $\kappa_t$ by the way of the simple relation $\kappa_t=k_0n_{co}\sin\theta$ and, as a consequence, the modes that will be excited and guided in the fiber. At the output of the fiber, one can image either the near-field (NF) intensity using a microscope objective or the far-field (FF) intensity in the focal plane of a suitable lens. 
Typical observations associated to a value of the transverse wave number $\kappa_t=36/R$ are presented in figure \ref{Speckle}. As diffraction occurs due to the finite aperture of the fiber, several transverse wavenumbers around the main transverse wavenumber $\kappa_t$ contribute so that more than one hundred modes are superposed at the output of the fiber. 
The NF figure \ref{Speckle}(a) shows a \textit{speckle} pattern: the intensity is statistically uniformly distributed over the whole cross section of the fiber. This figure is associated to the superposition of several modes each with a speckle-like behavior as predicted by M. V. Berry\cite{Berry1}. A nice experimental evidence of Berry's conjecture is given by the FF intensity figure \ref{Speckle}(b). In this figure, the intensity tends to fill a ring, whose mean radius is centered on the input transverse wavenumber $\kappa_t$ and its width is determined by the number of contributing modes. This illustrates that the field is constituted by the random superposition of a great number of plane waves each with a fixed transverse wave number but with random directions: the field of this generic so-called speckle mode can then be seen as a real Gaussian random variable.\\
The field that propagates along the fiber can be decomposed on the eigenmodes' basis $\{\phi_n\}$:
\begin{equation}
\psi(\textbf{r},z)=\sum_nc_n\phi_n(\textbf{r})e^{i\beta_nz}\label{summodes}
\end{equation}
where $c_n$ is the coupling coefficient between the input field $\psi_0(\textbf{r})$ and the eigenmode $\phi_n$, and $\beta_n$ is the propagation constant of the mode $\phi_n$ ($\beta$ being defined as $\beta^2=\beta_{co}^2-\kappa_t^2$). In the paraxial approximation, $\kappa_t\ll\beta_{co}$, and the exact expression of the decomposition of the field on the eigenmode's basis is: $\psi(\textbf{r},z)=\sum_nc_n\phi_n(\textbf{r})\exp({i(\beta_{co}-\kappa_t^2/(2\beta_{co}))z})$.\\
We should mention that our experimental system is an open cavity due to an absorbing cladding. It is now well established that open systems support non-real eigenfunctions\cite{Barthelemy,Savin}, whose complexness can be measured by the parameter $q^2$,
\begin{equation}
q^2=\frac{\langle (\Im m \phi)^2\rangle}{\langle (\Re e \phi)^2\rangle)}
\end{equation}
(with the condition that $\langle \Im m\phi \Re e \phi \rangle=0$). In Ref.\cite{Barthelemy,Savin}, it has been shown that $q$ can be related to the loss rate due to the leakage at the boundary. In the case of our fiber, measurements of attenuation along the fiber lead to q-values less than $10^{-2}$.\\
Thus, legitimely, the field of each mode can be seen as a real random Gaussian variable, one can easily deduce that the distribution of intensity of one mode $I=|\phi|^2$ follows a Porter-Thomas distribution. 
By superposing several modes, a Poisson distribution for the intensity of the field $\psi(\textbf{r},z)$ is then expected\cite{Doya0}. The figure \ref{PT} represents the probability distribution of intensity for the superposition of modes of the figure \ref{Speckle} and shows a good agreement with the Poisson prediction.
\\ 
We have experimentally investigated the series of scar modes associated to the shortest unstable periodic orbit that is the 2-bounce periodic orbit along the axis of symetry of the fiber's cross section. One can expect to observe a constructive interference effect building upon this periodic orbit if the following quantization condition for the value of the transverse wavenumber is fulfilled:
\begin{equation}
\kappa_t\mathcal{L}=2\pi p + \frac{\pi}{2} + 2\Delta \varphi\label{quantifscar}
\end{equation} 
where $\mathcal{L}$ is the length of the periodic orbit, $p$ an integer, $\pi/2$ is a phaseshift due to the existence of a self focal point on the 2-bounce periodic orbit, and $\Delta\varphi$ is the phaseshift due to reflection at the core/cladding interface. This phaseshift also depends on the value of the angle of the beam on the interface so that one has to solve the following transcendental equation to obtain the quantized value of the transverse wavenumber:
\begin{equation}
\kappa_t\mathcal{L}=2\pi p + \frac{\pi}{2} + 8\pi\arctan{\sqrt{\frac{n_{co}^2-n_{clad}^2}{k_t^2}\left(\frac{2\pi}{\lambda}\right)^2-1}}\label{eq_trans}
\end{equation}
 In the optical context, scar modes along the 2-bounce periodic orbit (PO) can be described as the modes of an unstable dielectric Fabry-Perot cavity.\\  
\begin{figure}[h]
\centering
\input{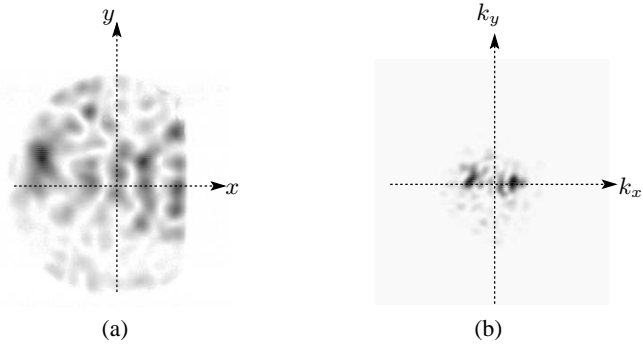}
\caption{Near-field intensity (a) and Far-field intensity (b) for a scar mode of order $p=5$ with $\kappa_t=10.35/R$.}
\label{Scar}       
\end{figure}\\
By solving the Eq. \ref{eq_trans}, one obtains quantized values of the transverse wavenumber associated to scar eigenmodes along the 2-bounce PO. 
For very precise illumination conditions, we have successfully excited a few modes around the scar mode of order $p=5$. The NF and FF intensity observed at the output of the fiber are presented in figure \ref{Scar}.\\
Even if several modes are excited, the intensity tends to localize along the direction of the 2-bounce PO emphasizing the main contribution of the scar mode in the superposition. In the FF figure, two symmetric peaks are observed and show that mainly one transverse wave number propagates along a well-defined direction, i.e., the direction of the 2-bounce PO. These observations have constituted the first experimental evidence of scar modes in an optical fiber\cite{Doya3}.\\
\section{Looking for optical scars}
\label{Scarsection}
The previous results have revealed the optical fiber as an ideal system to study the spatial behavior of waves in a chaotic billiard. If the properties of the field associated to a superposition of modes have been studied and analysed\cite{Doya2}, the individual eigenmodes cannot be simply observed due to diffraction at the fiber input. In particular, looking for an individual scar eigenmode appears to be a hard task. In order to investigate these rare eigenmodes, we have developped two approaches. The first one is dedicated to the signature of scar modes on the statistics of the field and intensity. Kaplan\cite{Kaplan1} and later Ishio\cite{Ishio} have studied the deviation from the Gaussian statistics due to the presence of scars. In section \ref{stat}, we present our preliminary results on the imprint of scar on the statistics. The second approach refers to the ability to perform a selective excitation of scar by the way of optical amplification. The problem of the introduction of a gain region in a system that has special properties of propagation also encounters the concerns of recent works on microlaser cavities or on disordered systems that will be made explicit in section \ref{amp}.
\subsection{Signature of \textit{scars} on the statistics}\label{stat}
\begin{figure}[ht]
\centering
\input{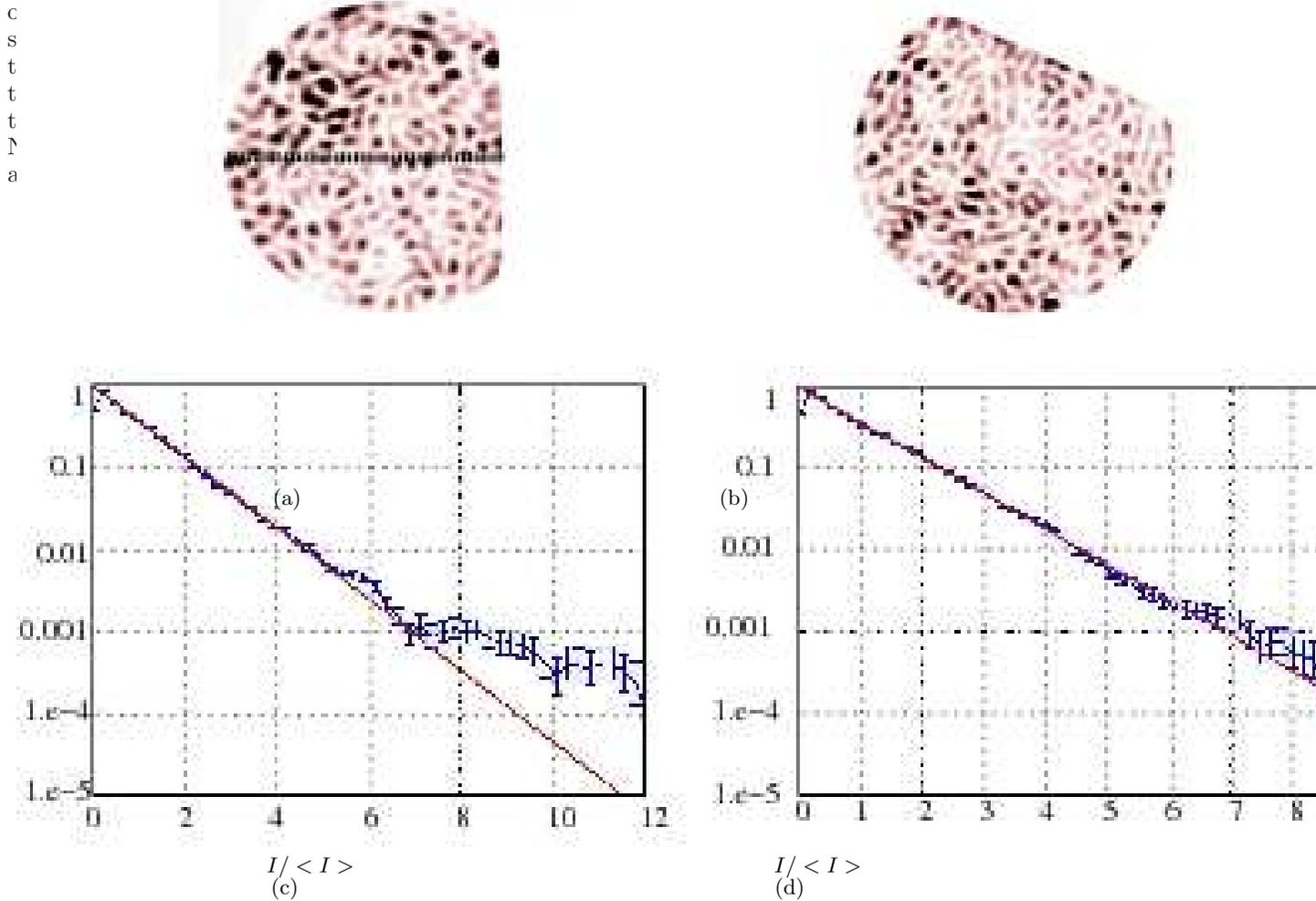}
\caption{NF intensities and their respective probability distributions for an input illumination along the 2-bounce PO (a), (c) and for the rotated fiber (b), (d).\label{manipsup}}
\end{figure}

The definition of \textit{scars} as an "\textit{extra-density [of quantum eigenstates] surround[ing] the region of periodic orbit}" in the pioneer paper by E. J. Heller\cite{Heller1} naturally leads to expect non standard properties for the distribution of the field's values. Deviations from Gaussian statistics deduced from Berry's conjecture for generic speckle modes have been studied numerically by L. Kaplan\cite{Kaplan1}. Using the plane wave decomposition method introduced by Heller, we have calculated the first 2000 modes of a D-shaped fiber with Dirichlet boundary conditions (that is a metal-coated D-shaped fiber). The behavior of these modes is quantitatively similar to those of our dielectric optical fiber as soon as we consider low order modes. The main differences will be due to the leakage of the modes in the dielectric fiber.\\ 
\begin{figure}[ht]
\centering
\input{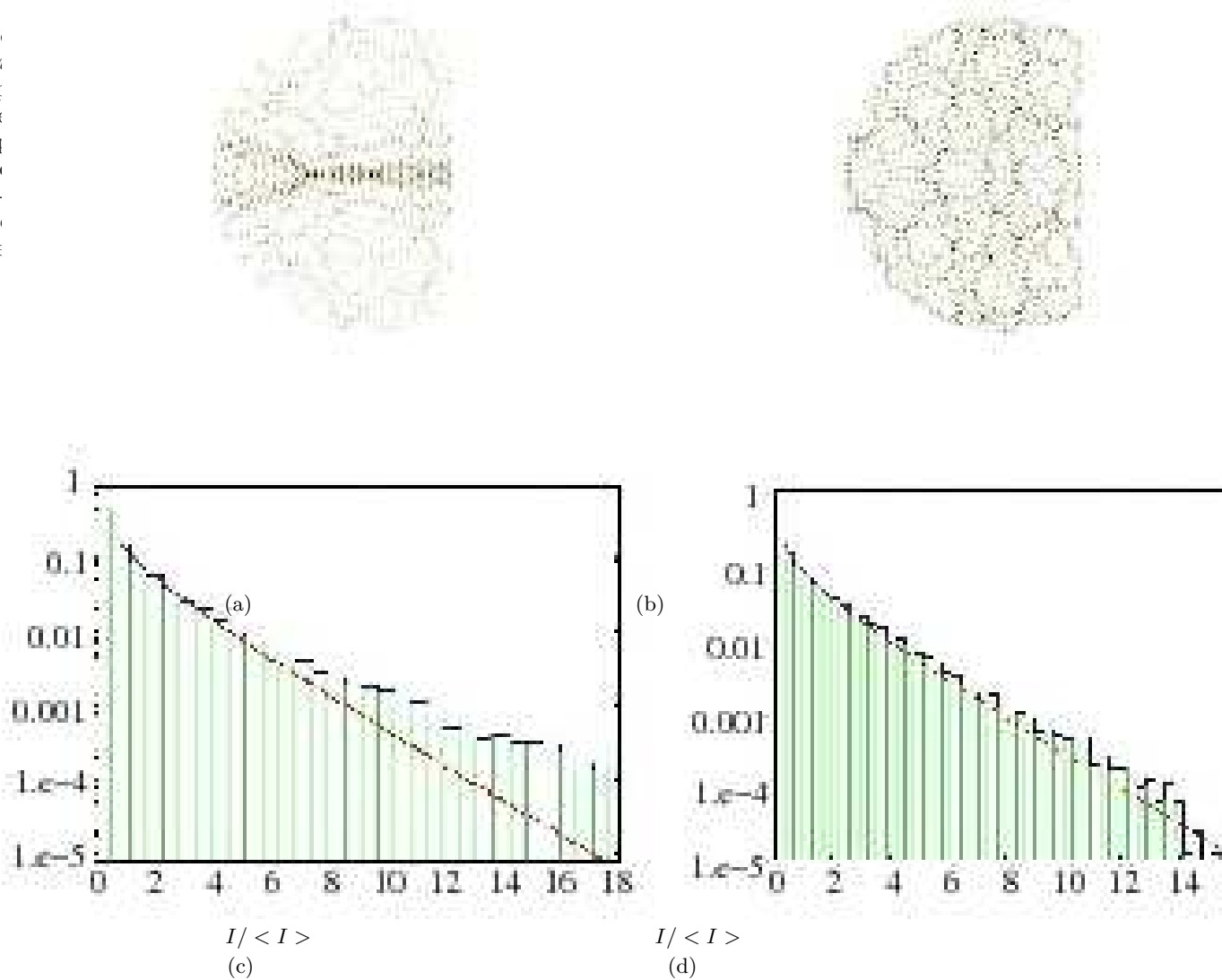}
\caption{Intensity of the field associated to calculated scar mode (a) and speckle mode (b) and their respective probability distributions of intensity (c) and (d).}\label{PWDMmodes}
\end{figure}\\
Figure \ref{PWDMmodes} displays the intensity of two calculated eigenmodes: the strong localisation of intensity along the 2-bounce PO observed for Fig.\ref{PWDMmodes}(a) corresponds to a scar mode, whereas Fig.\ref{PWDMmodes}(b) is associated to a generic speckle mode. 
The probability distributions of the intensity associated to these two characteristic modes compared to the Porter-Thomas distribution deduced from the Gaussian variable theory are shown in figure \ref{PWDMmodes}(c) et (d). If the probability distribution of the intensity for the speckle mode is in good agreement with the Porter-Thomas distribution, a strong deviation is observed for high values of intensity in the case of the scar mode. This deviation is related to the `extra-density' of the intensity along the direction of the periodic orbit: high values for scar modes are more probable than for speckle modes.\\
As scar modes seem to mark the intensity distribution, one can expect to look for the presence of scar modes in a superposition of several modes by analyzing the distribution of the intensity. This assumption has been studied in the context of open billiards by H. Ishio \textit{et. al.} who have revealed the influence of 'regular states' on the probability distribution of the superposition.\\
Experimentally, a favorable illumination is launched in the fiber in order to excite several modes around the scar mode. As the transverse wave number of the fiber equals $18.6/R$, the number of contributing modes is roughly 30 so that no scar mode seems to emerge in the NF intensity shown in Fig.\ref{manipsup}. Nevertheless, the probability distribution of intensity presents a strong deviation from the Poisson distribution for high intensity values. This deviation can be attributed to the presence of scar modes in the superposition. To check this assumption, we have turned the fiber so that the initial illumination was no longer in the direction of the 2-bounce PO. No manifestation of scar is detected neither in the NF nor in the intensity distribution that appears to be in very good agreement with the Poisson distribution. 
\\
Using the calculated eigenmodes, we have artificially constructed a superposition of five modes around a scar modes (see Eq.\ref{summodes}) where the coupling coefficients are fixed by the initial field $\psi_0(\textbf{r})$. In one case, the initial field is a plane wave with a transverse wave number associated to a scar launched in the direction of the 2-bounce PO. In the other case, the same plane wave is launched at an angle of 45 degrees. This launching condition is numerically equivalent to a rotation of the fiber around its axis in the experiment. In the latter case, the coupling coefficient of the scar mode is very small.
In both cases, we consider the field resulting from a 1-m propagation.\\
\begin{figure}[ht]
\centering
\input{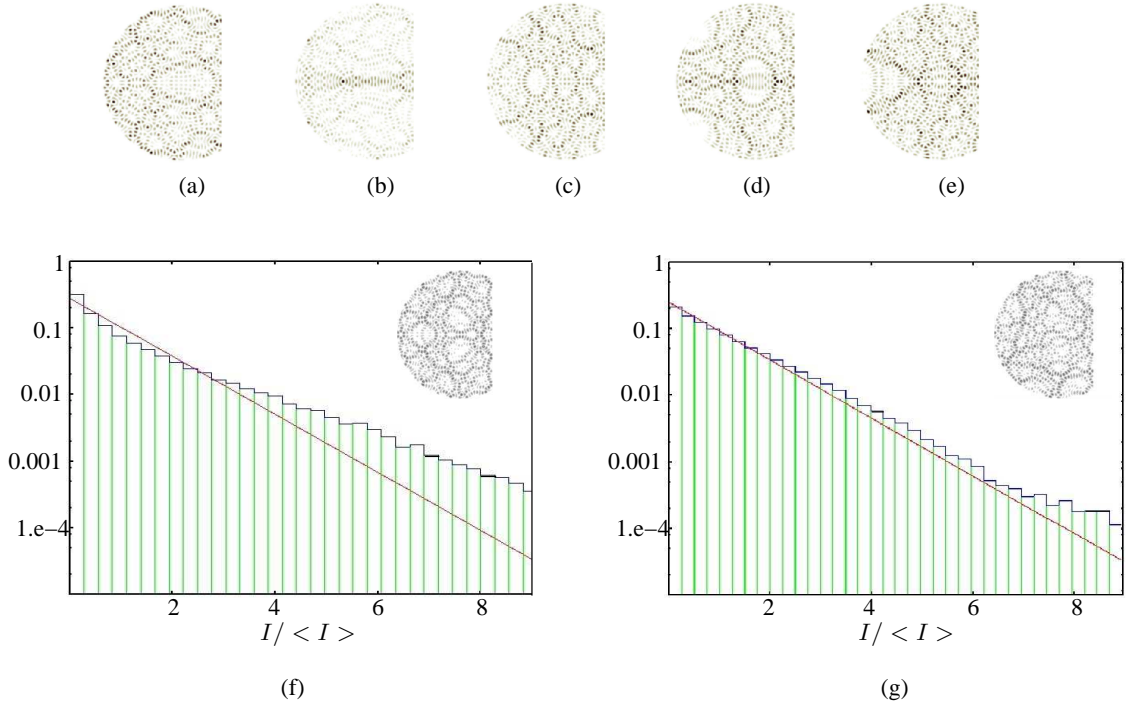}
\caption{Intensity of the field associated to a scar (b) and its four neighouring speckle modes (a), (c), (d), (e). Probability distribution of intensity for a superposition of these modes (shown in the inset) associated to an input illumination in the direction of the scar (f) or an arbitrary input illumination (g).\label{supmodes}}
\end{figure}\\
The figure \ref{supmodes} presents the NF intensity and the probability distribution of the intensity for the two different launching conditions. The NF intensity is statistically uniformly distributed over the whole transverse cross section showing no difference between the two cases. Nevertheless, a deviation for high value of intensity is also observed for the ``scar" illumination whereas the Poisson distribution is observed for the rotated initial condition.   \\
As expected, as no scar is excited, no evidence of scar is observed. These results are preliminary and will be completed by a more systematic experimental investigation in a forthcoming publication. 
\subsection{Optical Amplification of Scars}
\label{amp}
\subsubsection{General context}\label{ampcontext}
Kudrolli \textit{et al.} have studied the influence of nonlinear Faraday waves on the selection of scar modes in a stadium shape container\cite{Kudrolli01}, but only very few investigations have been done on the behavior of the chaotic wavefunctions in a nonlinear system. Recently, chaotic microlasers associated to wave chaos in the presence of gain have been the support of a great number of investigations\cite{Gmachl,Harayama03}. These studies have shown that a lasing effect occurs for scar modes thanks to their high quality factor, and the privileged directions of emission associated to scar modes appear to be interesting to provide highly directional lasers.\\In the context of wave propagation in complex media, some of the spatial properties of waves in a disordered system are somewhat similar to those met in chaotic systems.  For instance, numerical investigations have recently shown that the amplified modes in a random laser are the modes of the passive disordered system \cite{Sebbah_PRB_2002}. The same authors have shown  that the identification  between lasing modes and passive modes holds even when the gain region is smaller than the mode size\cite{Vanneste_PRL_2001}.

We have shown that a passive optical fiber with a D-shape constitutes a powerful experimental system for studying wave chaos. But above all, multimode D-shaped fibers are also widely used to achieve optical power amplifiers \cite{Desurvire} in which a telecom signal (called the signal) is amplified at the cost of an ancillary signal (called the pump). Usually the optical index of the Erbium or Ytterbium doped-core is higher than the index of the surrounding D-shape cladding. The signal propagates in a single mode doped core, and, by guiding the pump, the multimode fiber only plays the role of an energy reservoir. By some appropriate modifications of the fabrication process, it is possible to obtain a negligible index mismatch between the small active region and the multimode fiber \footnote{The actual fiber has been fabricated in our institute by W. Blanc, in collaboration with P. Oupicky from Plasma Institute of Physics in Turnov (Czech Rep.)}, at least sufficiently low to ensure that no guided mode exists inside the doped area. Then, the signal and the pump propagate on the modes of the entire section of the multimode D-shape fiber. In the following, we present  realistic numerical simulations of such a chaotic fiber with an Ytterbium-doped region. The simulation is based on the Beam Propagation Method we already used successfully for simulating optical amplification \cite{Doya_OptLett}.  We show how localized gain may selectively amplify a family of scar modes.
\\
\subsubsection{Results}\label{ampresults}
\begin{figure}[h]
\centering
\input{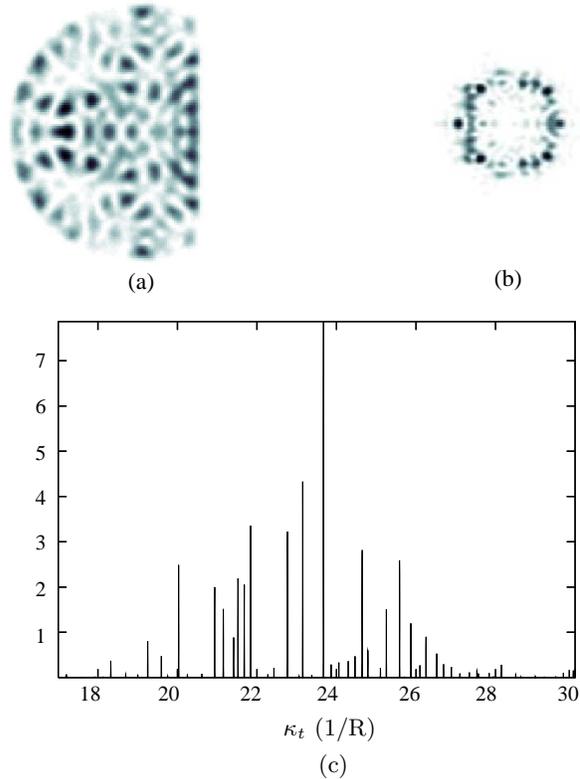}
\caption {NF intensity (a) and FF intensity (b) at the output of the passive fiber for a plane wave illumination in the direction of the 2-bounce scar with $\kappa_t=23.45(1/R)$ and the pseudo-spectrum (c).\label{PWPassive}}
\end{figure}

\begin{figure}[ht]
\centering
\input{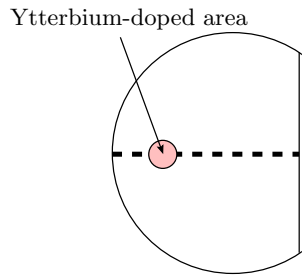}
\caption{Schematic view of the transverse crosssection of the active fiber with an Ytterbium doped area.\label{dopedfiber}}
\end{figure}

In our multimode fiber, selective excitation of one eigenmode appears to be a tremendous task. Selective excitation techniques as holographic masks of the modes or intracavity filtering are not appropriate in our case due to the complex patterns of the modes of the chaotic fiber. An other technique of selective excitation consists in  controlling the transverse wavenumber that propagates along the fiber. This can be done by tuning the angle of a plane wave illumination at the input of the fiber. Nevertheless, due to the finite aperture of the fiber, diffraction occurs leading to a broadening of the value of the transverse wavenumber that propagates along the fiber. For a plane wave $\exp(-i\kappa_tx)$ launched along the direction of the 2-bounce PO with a value of $\kappa_t=23.45/R$, we evaluate the spectral broadening to $\delta \kappa_t=14/R$. As a consequence, almost 50 modes are excited preventing any possibility to maintain the modal structure of the scar along the propagation. To identify the guided modes along the propagation, we calculate what we call a pseudo-spectrum of our system. This pseudo-spectrum is derived from a standard semiclassical procedure developped by Heller\cite{TOM} and adapted to our numerical algorithm \cite{FEI}. The pseudo-spectrum is extracted from the Fourier transform of the correlation function $\mathrm{C}(z)$ obtained from the overlap of the initial field $\psi_0(\textbf{r})$ with the propagating field $\psi(\textbf{r},z)$. The pseudo-spectrum of a plane wave illumination $\exp(-i\kappa_tx)$ with $\kappa_t=23.45/R$ together with the NF and FF intensity at the output of the fiber are plotted in figure (\ref{PWPassive}).  
In the NF figure, the intensity is statistically uniformly distributed over the transverse cross section of the fiber showing no localisation along the 2-bounce PO associated to the expected scar mode. The loss of information about the direction of the initial  plane wave can clearly be seen in the FF intensity figure where the light tends to fill a ring whose radius is centered on the input transverse wavenumber $\kappa_t$. These pictures are actually associated to a superposition of the roughly 50 modes initially excited as can be seen in the pseudo spectum centered on the value of the transverse wavenumber associated to the scar. 
\\
Introducing doping elements in an optical fiber is a standard process to obtain optical amplification for instance. As the scar modes are characterized by an enhancement of intensity along some particular direction in the transverse cross section of the fiber, we introduce  active ions along that direction to perform a selective amplification through the strong interaction of the field with the doped area. With the experiment in mind, we have numerically investigated realistic systems. What can be done experimentally is to introduce active rare earth ions in a circular area centered on the axis of the 2-bounce PO. To make the doping area efficient for our purpose, two conditions have to be fulfilled: a)the location of this area has to be optimized to favor the amplification of the 2-bounce PO scar, b) this area should have the smallest feasible refractive index difference with the core refractive index to prevent light from being guided in the doped area. As a consequence, the doped area is located close to the position of the self-focal point of the 2-bounce PO that also corresponds to a maximum of intensity of the scar modes (see figure \ref{dopedfiber}). Experimentally, the introduction of 1000ppm of Ytterbium ions leads to an increase of the refractive index equal to $2.10^{-4}$. In spite of this small refractive index difference between the doped area and the D-shaped core, light is actually guided in the doped area and amplified. To overcome this problem, we use a Gaussian refractive index profile which seems to support no guided modes. \\
\\
In this kind of system, optical amplification occurs due to the transfer of energy from an optical pump wave at the wavelength, $\lambda_p=980$nm in our case, to the signal to be amplified at the wavelength $\lambda_s=1020$nm. Ytterbium is a four-level system that permits to use different wavelengths for the pump and the signal preventing direct signal reabsorption. As a consequence, Ytterbium in a silica matrix can be treated as a three energy-level system with a metastable third energy level.  

\begin{figure}[h]
\centering
\input{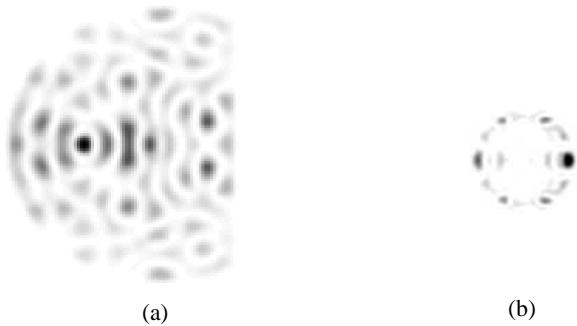}
\caption { NF intensity (a) and  FF intensity (b) of the signal at the output of the active fiber for a launching initial condition in the direction of the scar with $\kappa_t=23.45/R$.\label{PWActive}}
\end{figure}

\begin{figure}[h]
\centering
\input{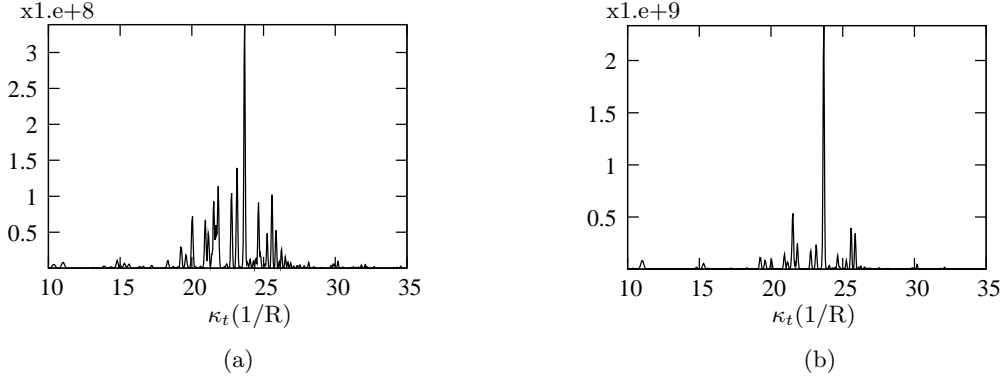}
\caption {Evolution of the intensity spectrum along the propagation (a) at z=0cm, (b) z=80cm .\label{ActiveSpectrum}}
\end{figure}

Using the evolution equations of the population density, one can easily obtain the following expressions for the evolution of the pump and signal intensities\cite{THY}:
\begin{eqnarray}
\frac{dI_p}{dz} &=& -\sigma_{pa}N_1(z)I_p(z)\label{eq-pump}\\
\frac{dI_s}{dz} &=& \sigma_{sa}(\eta_sN_2-N_1)I_s(z)
\label{eq-signal}
\end{eqnarray}
where $I_p$ and $I_s$ are the intensity of the pump and of the signal respectively, $\sigma_{sa}$ the absorption cross section of the signal, $\sigma_{pa}$ the absorption cross section for the pump and $\eta=\sigma_{se}/\sigma_{sa}$ with $\sigma_{se}$ the emission cross section of the signal. As soon as the intensity of the signal $I_s$ is low compared to the intensity of the pump $I_p$, Eq. \ref{eq-pump} only depends on $I_p$ and the two equations are decoupled so that they can be treated separately. This case is fulfilled in our simulation where the initial signal power is $P_{s_0}=100\mu W$ and the pump power equals to $P_{p_0}=3W$.\\
The \textsc{Bpm} algorithm is modified to take equations \ref{eq-pump},\ref{eq-signal} into account to simulate the coupled evolutions of the pump and the signal. To make an efficient transfer of energy from the pump to the signal, one has to maximize the number of excited pump modes: a focused laser beam at the pump wavelength is therefore launched in the multimode core. A great number of modes of the D-shaped core are thus excited and the overlap of their fields with the doped area is responsible for the energy transfer from the pump to the signal. In previous works, we have shown that the chaotic D-shaped geometry is an efficient shape for this purpose\cite{LEP}. The evolution of the pump along the fiber follows an early exponential decay due to absorption in the doped area that tends to saturate as soon as the pump power is no longer able to fulfill the population inversion condition $\eta_s N_2-N_1>0$.\\
We are now interested in the evolution of the signal associated to a plane wave illumination in the direction  of the 2-bounce PO of the selected scar with $\kappa_t=23.45/R$. In figure \ref{PWActive}, we plot the NF and FF intensities computed for a distance  $z_{max}\approx 2$m associated to the maximum gain value.  
\\
The intensity figures clearly show that the 2-bounce PO direction is privileged: in the NF intensity, high intensities concentrate on the horizontal direction associated to the scar, and in the FF intensity, two symmetric spots of intensity arise, located on the value of the initial $\kappa_t$. The assumption, that the expected scar mode seems to emerge thanks to amplification, is confirmed by the pseudo-spectrum. The figure \ref{ActiveSpectrum} represents the evolution of the pseudo-spectrum along the propagation. It is deduced from pseudo-instantaneous spectra computed for three sample of propagation of length $5$cm along the 4-m fiber. From a broad spectrum for $z=0$m, we observe that a selection of few peaks tends to emerge. Only some modes are being amplified along the propagation, and the mode associated to the scar with value of $\kappa_t=23.45/R$ distinctly dominates the spectrum.\\ 
\\
\begin{figure}[h]
\centering
\input{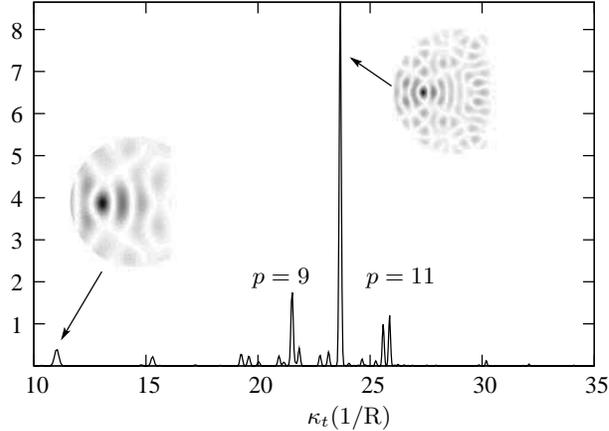}
\caption {Spectrum calculated around the maximal value of the gain and the calculated modes associated to its main peaks.\label{filtering}}
\end{figure}\\
To clearly identify the excited modes, we perform a filtering on the FF by applying a ring with mean radius $\kappa_t$ and a Gaussian profile, $\kappa_t$ being associated to one of the main peaks of the spectrum. The corresponding NF is deduced from the inverse Fourier transform. By applying the previous procedure to the value of $\kappa_t$ related to the scar of order $p$=10, we have checked that the spatial distribution of the intensity is associated to the corresponding scar mode (fig \ref{filtering}). We have also checked that the peak labelled by $\kappa_t=10.9/R$  corresponds to the intensity pattern of the scar of order $p$=4 along the 2-bounce PO. The NF intensities obtained by applying the filtering procedure on different values of $\kappa_t$ respectively associated to the scar modes of order $p$=4 and $p$=10 are plotted in an insert on figure \ref{filtering} and compared to the corresponding modes calculated for a waveguide with Dirichlet conditions. The  intensity patterns for the two filtered-NF are similar to the individual calculated scars: the scar modes are amplified to detriment of the ergodic modes. \\
To check the critical effect of the location of the gain region on scar amplification, we have changed the location of the doped area. By moving the location of the doped region away from the self-focal point of the 2-bounce PO, we observe no scar amplification confirming that the gain region must have a maximal overlap with the scar to be efficient.\\

 The localization of light along the 2-bounce PO clearly demonstrates that the scar modes have been excited and enhanced thanks to the interaction with the doped region.
\section{Conclusion}
The optical fiber with a truncated cross section appears to be a good candidate to image experimentally wave chaos: complementary informations about the field that propagates in the chaotic billiard are deduced from the NF and FF intensity. The ergodic behavior of generic modes has been proved experimentally and scar modes have first been observed in an optical fiber. Due to the inescapable diffraction at the fiber input, the properties of individual scar modes cannot be studied experimentally. Nevertheless, their signature on the probability distribution of the intensity appears to be pregnant: our preliminary investigations have shown that deviations to the ergodic theory are to be expected as soon as a scar is excited on a superposition of modes. \\To perform a selective excitation of scar modes in our multimode structure, we have investigated the influence of a localized gain area on the propagation of a plane-wave illumination. Our numerical results clearly show that scar modes are selectively amplified. The experimental validation is in progress and will be the subject of a next publication. 
\subsection*{Acknowledgments}
We would like to thank the \textit{Active Optical Fiber} group of the \textsc{Lpmc} for the realization of our ``exotic" preform and the undergraduate and master students A. Martin and S. Onteniente for their contribution to this work. This research is part of the project \textit{Chaos On Line} $n^o$ANR-05-JCJC-0099-01 supported by the National Agency of Research (\textsc{Anr}).



\end{document}